# The new SOXS instrument for the ESO NTT

P. Schipani*[a], R. Claudi[b], S. Campana[c], A. Baruffolo[b], S. Basa[d], S. Basso[c], E. Cappellaro[b], E. Cascone[a], R. Cosentino[e,f], F. D'Alessio[g], V. De Caprio[a], M. Della Valle[a], A. de Ugarte Postigo[h], S. D'Orsi[a], R. Franzen[i], J. Fynbo[j], A. Gal-Yam[k], D. Gardiol[l], E. Giro[b], M. Hamuy[m], M. Iuzzolino[a], D. Loreggia[l], S. Mattila[n], M. Munari[f], G. Pignata[o], M. Riva[c], S. Savarese[a], B. Schmidt[i], S. Scuderi[f], S. Smartt[p], F. Vitali[g]

[a]INAF - Osservatorio Astronomico di Capodimonte, Salita Moiariello 16, I-80131, Napoli, Italy
[b]INAF – Osservatorio Astronomico di Padova, Vicolo dell'Osservatorio 5, I-35122 Padova, Italy
[c]INAF – Osservatorio Astronomico di Brera, Via Bianchi 46, I-23807 Merate (LC), Italy
[d]Aix Marseille Université, CNRS, LAM (Laboratoire d'Astrophysique de Marseille) UMR 7326, 13388, Marseille, France
[e]FGG-INAF, TNG, Rambla J.A. Fernández Pérez 7, E-38712 Breña Baja (TF), Spain
[f]INAF – Osservatorio Astronomico di Catania, Via S. Sofia 78 30, I-95123 Catania, Italy
[g]INAF – Osservatorio Astronomico di Roma, Via Frascati 33, I-00078 Monte Porzio Catone, Italy
[h]Instituto de Astrofisica de Andalucia, Glorieta de la Astronomía, E-18008 Granada, Spain
[i]ANU RSAA – Mount Stromlo Observatory, Cotter Road, Weston Creek, ACT 2611, Australia
[j]DARK Cosmology Center, Juliane Maries Vej 30, 2100 Copenhagen, Denmark
[k]Weizmann Institute of Science, Herzl St 234, Rehovot, 7610001, Israel
[l]INAF – Osservatorio Astrofisico di Torino, Via Osservatorio 20, I-10025 Pino Torinese (TO), Italy
[m]Universidad de Chile (DAS) / MAS, Camino El Observatorio 1515, Las Condes, Santiago, Chile
[n]FINCA - Finnish Centre for Astronomy with ESO, 20014 Turun yliopisto, Finland
[o]Universidad Andres Bello, Avda. Republica 252, Santiago, Chile
[p]Astrophysics Research Centre, Queen's University Belfast, Belfast, County Antrim, BT7 1NN, UK

## ABSTRACT

SOXS (Son Of X-Shooter) will be a unique spectroscopic facility for the ESO-NTT 3.5-m telescope in La Silla (Chile), able to cover the optical/NIR band (350-1750 nm). The design foresees a high-efficiency spectrograph with a resolution-slit product of ~4,500, capable of simultaneously observing the complete spectral range 350 - 1750 nm with a good sensitivity, with light imaging capabilities in the visible band. This paper outlines the status of the project.

**Keywords:** Spectrographs, Instrumentation, Transients

## 1. INTRODUCTION

SOXS is a spectroscopic facility for the follow-up of transient sources, proposed by an Italy-led international consortium (PI: S. Campana). The instrument has been selected in 2015 out of 19 proposals, in response to a competitive call for scientific ideas at the NTT issued by ESO in 2014, which went through two down-selection steps. It is supposed to start operations in early 2020s, replacing the current NTT instrumentation.

We foresee a unique spectroscopic facility able to cover the optical/NIR band (0.35-1.75 μm). The acronym of the instrument originates from its similarities with X-Shoother [1]. SOXS will also benefit of the example of the NOT Transient Explorer (NTE) at the Nordic Optical Telescope (2.5-m), a parallel project but for the northern sky. SOXS will be a medium resolution spectrometer (R~4,500). The limiting magnitude of *R*~20 (1 hr at S/N~10) is perfectly suited to study transients from on-going and future imaging surveys. Light imaging capabilities in the optical are also foreseen to allow for multi-band photometry of the faintest transients, with a field of view of at least 2 arcmin.

*pietro.schipani@oacn.inaf.it

The consortium will provide observers in Chile for extended periods and will carry out rapid and long term Target of Opportunity (ToO) requests, similarly to what is now occurring with PESSTO [2], but on a variety of astronomical objects. SOXS will be fed by all kind of transients from different surveys. These will be a mixture of fast alerts (e.g. gamma-ray bursts, gravitational waves), mid-term alerts (e.g. supernovae, X-ray transients), fixed time events (e.g. close-by passage of a minor body), etc.

The observing schedule will have to be updated daily and flexible to accommodate fast alerts. We will dedicate part (~5%) of the SOXS observing time to fast ToO open to the community, in order to broaden even more our science case. Raw data will be immediately public to the entire community and we intend to have staged releases of calibrated data.

## 2. SCIENCE CASE

In the near future, we will enter the golden age of time-domain astronomy having in place deep ground-based optical surveys (Zwicky Transient Factory, PanSTARRS, Dark Energy Survey, La Silla Quest, SkyMapper, VST, VISTA, Large Synoptic Survey Telescope, etc.), space-based optical surveys (Gaia and Euclid), high-energy instruments (Swift, Fermi, INTEGRAL, MAXI and SVOM), radio surveys (LOFAR, SKA), Gravitational Wave (GW) experiments (A-Virgo, A-Ligo), neutrino experiments (KM3Net), all calling for a rapid follow-up and characterisation (and distance determination) of the detected transients. A bibliographic search showed that among astronomical papers published on Nature in the 2005-2014 time frame, ~40% of them are on transient objects/events. The discovery space in this research field is potentially immense, including virtually all astronomy disciplines. A dedicated suitable spectroscopic facility able to exploit the science of these transients is lacking, resulting in severe science "dissipation".

The science case is then very broad, given the versatility of such an instrument and the intrinsic variability of the sky at all wavelengths. It ranges from moving minor bodies in the solar system, to bursting young stellar objects, cataclysmic variables and X-ray binary transients in our Galaxy, supernovae and tidal disruption events in the local Universe, up to gamma-ray bursts in the very distant and young Universe, basically encompassing all distance scales and astronomy branches. Although the focus is on transients and variables, there is anyway a wide range of astrophysical targets and science topics that could benefit from an instrument with the capability of SOXS at the NTT.

A spectroscopic facility to follow-up transient sources is a unique tool to provide a service to the community at large and to do great science. Time-domain optical surveys are revolutionising the optical astronomy providing insight in basically all areas. Transients discovered by current surveys have limiting magnitudes $R\sim20$, that are well suited to be followed-up by a 4 m-class telescope. High-energy missions (e.g. Swift, INTEGRAL, Fermi) are detecting in real-time transients sources calling for a fast follow-up. SOXS will be a key instrument to provide the spectroscopic partner to any kind of transient survey. The experience of the Swift high-energy mission in the study of the X-ray sky is enlightening. A fraction of the Swift observing time is fully open to Target of Opportunity (ToO) observations from the community. A great use of this time has been done since now, considerably contributing to the success of the Swift mission. In the optical, especially concerning the spectroscopic domain, a similar possibility of fast triggering short observations is lacking (unless very highly ranked pre-approved programs adopting a Rapid Response Mode).

## 3. INSTRUMENT DESIGN

The instrument will be located at one of the two Nasmyth interfaces of the NTT. The design of the skeleton structure and the arrangement of the spectrographs are such that SOXS well fits within the volume specified for the Nasmyth focus of the NTT telescope. The design is optimised for low torque at the telescope flange by keeping the centre of gravity as close as possible to the telescope focus.

The current baseline design consists of a central structure, the common path (CP), which supports two prism cross-dispersed echelle spectrographs optimised for the UV-visible and near-IR wavelength ranges. Attached to the CP are the necessary calibration and acquisition facilities, dichroics and relay optics.

The instrument can be broken down into the sub-systems listed below.

- **Common Path**: the backbone of the instrument and the interface to the NTT Nasmyth focus flange. The light coming from the focus of the telescope is split by the common optics into two different optical paths in order to feed the two spectrographs. Both the A&G camera and the calibration unit are connected to the two scientific arms through the common path.
- **UV-VIS Spectrograph**: an echelle-dispersed spectrograph working in the wavelength range 350-760 nm.

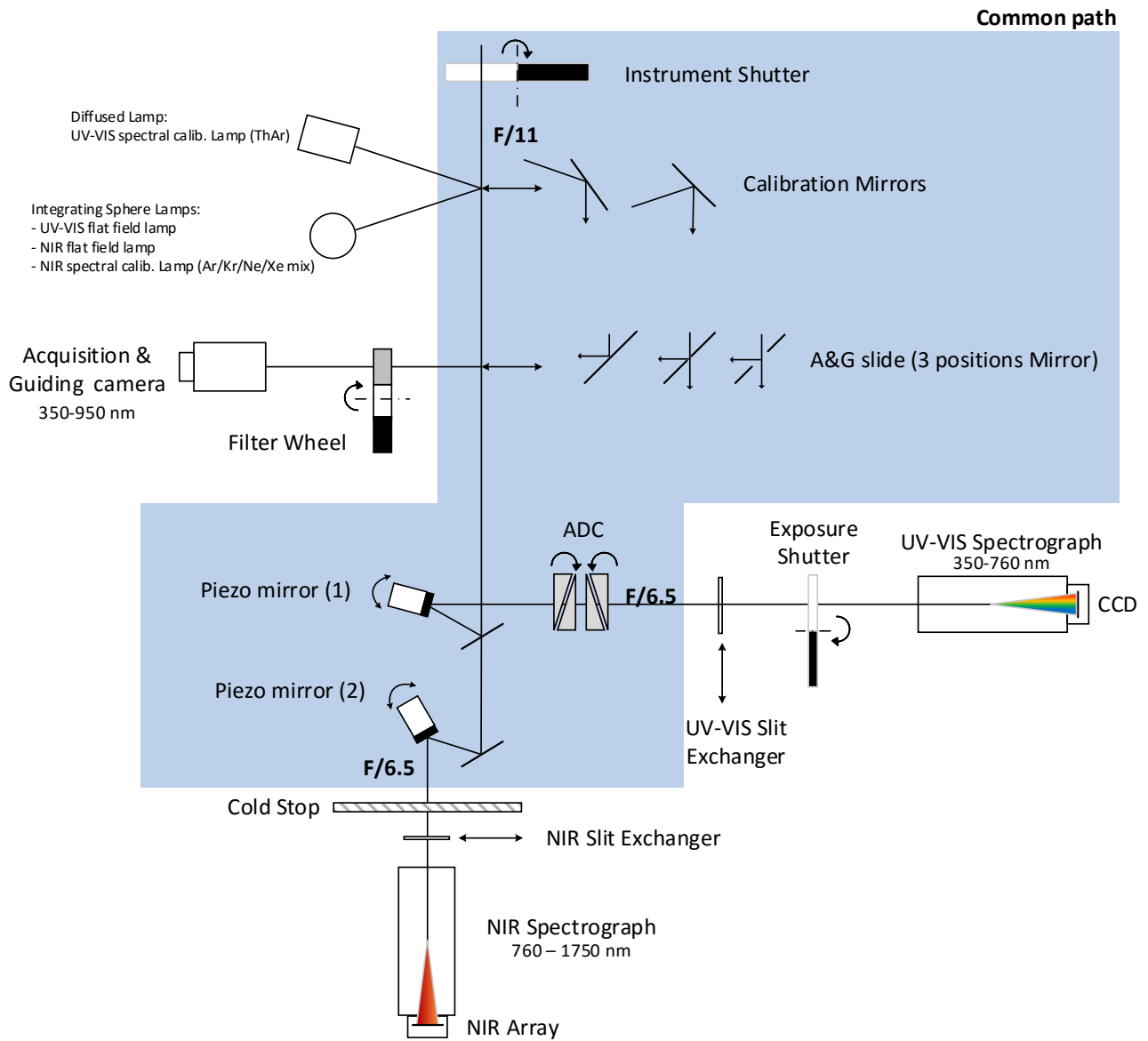

Figure 1. SOXS Schematic diagram.

- **NIR Spectrograph**: an echelle-dispersed spectrograph working in the wavelength range 760-1750 nm, being the 760 nm cut wavelength coincident with a prominent atmospheric feature.
- **UV-VIS Detector System**: it includes the detector, the front-end electronics, the mechanical support, the controller and the optics for the UV-VIS arm. It interfaces with the cryogenic sub-system.
- **NIR Detector System**: it includes the array detector, the front-end electronics, the mechanical support, the controller and the optics for the NIR arm. It interfaces with the cryogenic sub-system.
- **A&G Camera**: acquisition and guiding camera. It allows for the centring of the point source on the selected slit and makes a second order guiding available. This sub-system might also be used as light imager in a given photometric band in order to perform photometry and flux calibration. A set of standard optical filters will be adopted (e.g. SDSS).
- **Calibration Unit**: it includes all the lamps for both wavelength and photometric calibration and all the necessary optics to perform calibrations.

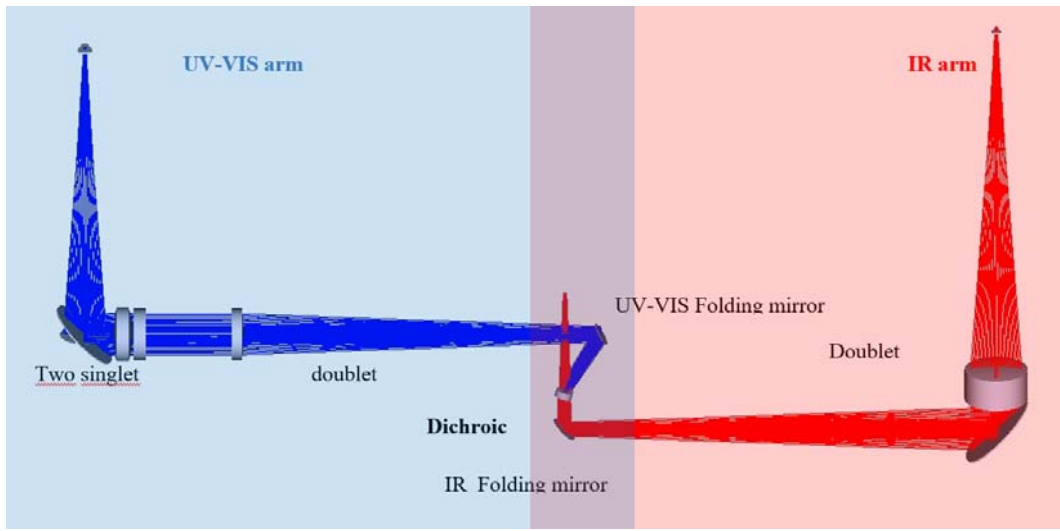

Figure 2. Common path optical layout.

- **Cryogenic System**: all the hardware and electronics for the housekeeping of cryogenic functions for both the AIT and the operational activities. Both the dewars of the UV-VIS and NIR detector are within this sub-system.
- **System Electronics and Control**: all the control functions including electric, electronics and control interfaces for all the sub-systems.
- **Instrument Software**: the software that allows the operation and maintenance of the instrument.
- **Data Reduction and Handling**: all the recipes needed to reduce the rough spectra produced by the instrument.

The conceptual arrangement is shown in Figure 1.

## 4. COMMON PATH

The SOXS Common Path has the following functions:
- relay the telescope focus on the two spectrographs slit planes
- modify the F/# of the telescope
- match the entrance pupil of the spectrographs with the exit pupil of the CP
- split the light according to wavelength, separating it in two arms, each feeding one of the spectrographs
- host the Atmospheric Dispersion Corrector (ADC) for the UV-VIS Arm
- form a cold stop in the IR arm

The layout of the common path optics in the current baseline is shown in Figure 2.
The light coming from the telescope focus is split by a dichroic according to the wavelength, reflecting the "blue" (<760 nm) light and transmitting the "red" light (>760 nm). The blue light is reflected on the first surface of the dichroic, to avoid problems of transmission of short wavelengths due the substrate. Moreover, the incidence occurs at a small angle (15°) with respect to normal, in order to ensure the steepest possible cut in wavelength.
After the dichroic the light follows two different paths.
The "blue" light is reflected again by a folding mirror (that can be used for flexure compensation, if needed) in a direction parallel to the plane of the Nasmyth focus, going through the refractive optical group that reduces the F/# and compensates for the atmospheric dispersion, using counter-rotating prisms. The first optical element is a cemented doublet (first surface aspheric), forming a collimated beam where the Atmospheric Dispersion Corrector is placed. It is composed by a couple of two counter-rotating glass prisms. A second air separated doublet (all spherical elements) and a singlet relay the focal plane onto the entrance slit of the UV-VIS spectrograph, and a field lens matches the pupils.
The "red" light, transmitted by the dichroic, is reflected by a 45° mirror (which, again, can be used for flexure compensation if needed) in the opposite direction of the blue one, and then toward the IR spectrograph, and is refocused with a reduced F/# by a doublet (first surface aspheric). A field lens matches the pupils.
In both arms, the glasses have been chosen to optimise the throughput and, obviously, the optical performances.

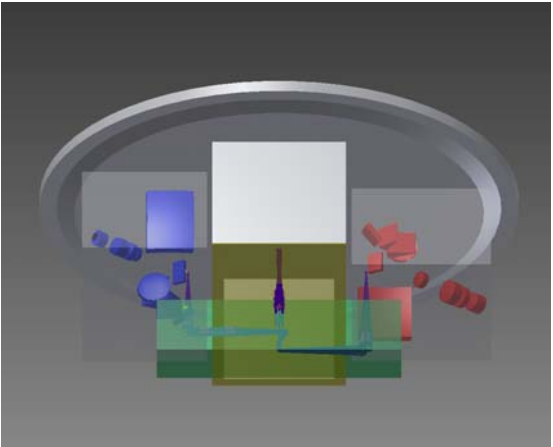
Figure 3. Architectural concept.

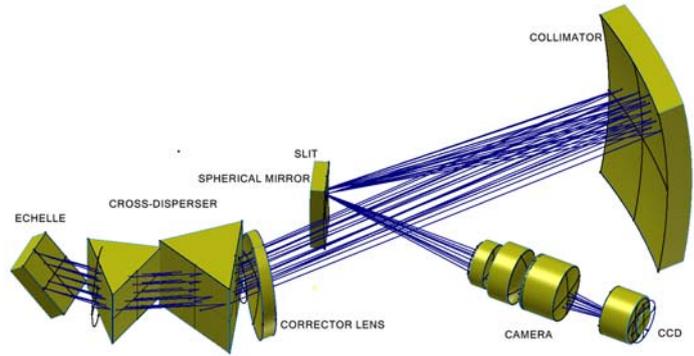
Figure 4. 4C layout.

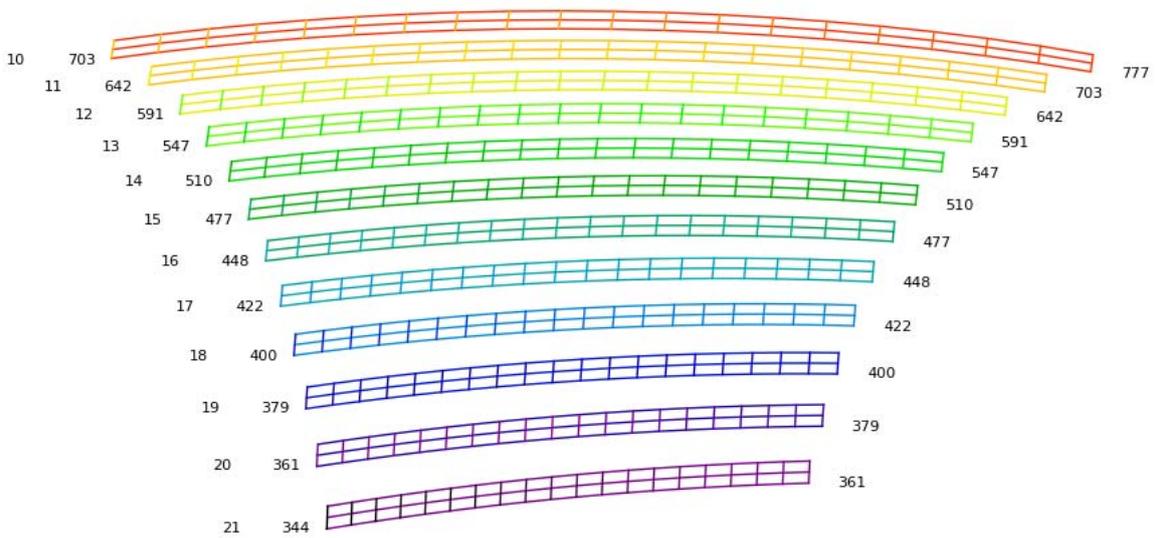
Figure 5. UV-VIS spectrograph spectral order format.

The skeleton structure of the SOXS instrument will be designed, for the sake of the stiffness to weight ratio optimisation, in aluminium with the two spectrographs mounted at each side (Figure 3). All the pre-slit functions will be allocated in a self-supported box fixed on the bottom part of the skeleton structure. The skeleton structure will be mounted on the NTT telescope rotator flange. The field electronics will be arranged at the free sides of the common path structure.

The present design of the common path mechanics, the arrangement of the two spectrographs and the pre-slit box are such that SOXS fits within the volume of the space envelope of the Nasmyth focus B at the NTT.

## 5. UV-VIS AND NIR SPECTROGRAPHS

The UV-VIS spectrograph of SOXS is based on the 4C (Collimator Compensation of Camera Chromatism) concept [3] (Figure 4). The light coming from the slit is collimated by an off-axis Maksutov-type collimator (spherical mirror and spherical corrector lens), then goes through the cross-disperser and impinges on a grating, acting as main disperser. It then goes back through the cross-disperser and the collimator again, forming a first echelle spectrum on a plane near the entrance slit. This spectrum is then relayed, by a pupil mirror and a camera, onto the detector.

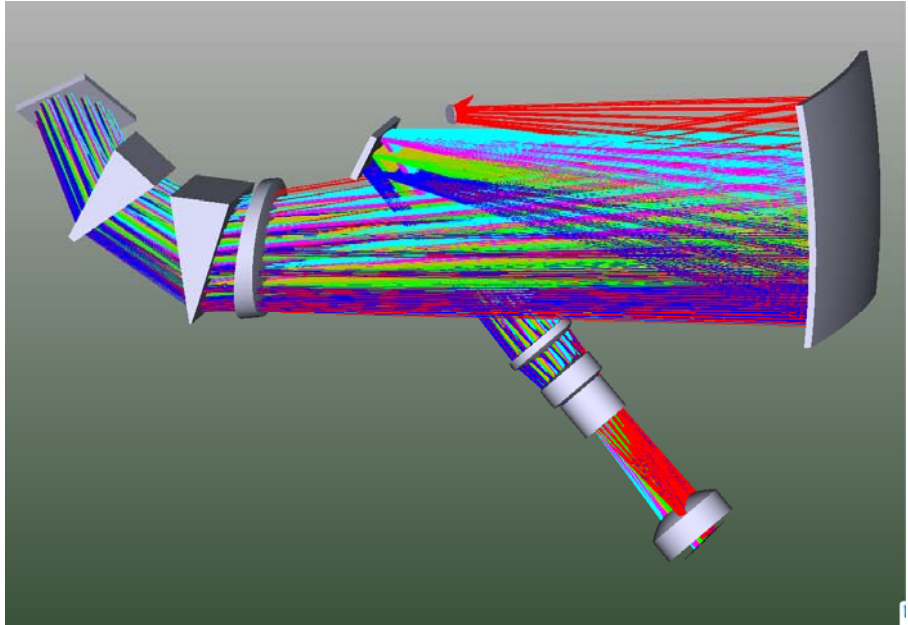

Figure 6. NIR spectrograph layout.

The collimator has an entrance F/# matching the exit one of the common path; the spherical mirror is used in an area corresponding roughly to a rectangle 180x254 mm (even if it will be possible to cut it in a smaller shaper), to reduce the weight; the corrector lens (diameter 110 mm) is made in Silica, to allow for the maximum throughput in UV.
For the same reasons, the cross-disperser prisms are in the same material. Two prisms are foreseen to get enough cross-dispersion of the spectrum from 360 nm up to 760 nm. The grating considered so far has a blaze angle of 41.77° and 180 grooves/mm; it is used at Littrow angle, and with a small off-plane angle (3.10° in the baseline design). The dimension of the (monochromatic) beam on the grating is ~45 mm width, resulting in the desired resolution (~5,000 for a 1 arcsec slit). The physical dimension of the grating is grater then 45 mm, to permit the accommodation of beams pre-dispersed by prisms. We foresee to use the orders from 10 to 21.
The camera, after a spherical mirror that recreates a pupil, is composed by 4 elements (one doublet), with one aspherical surface (first surface of the first lens and first lens of the doublet). The detector is slightly tilted (~0.74°). The selected glasses ($CaF_2$, and PBL25Y) have been selected to guarantee a high throughput down to 350-360 nm in the UV.
The spectral format obtained on the detector is shown in Figure 5; it fits into a 2k x 2k, 15 μm pixel detector. The orders are shown with their number and the free spectral range extreme wavelengths, in nanometers. The minimum separation between the orders occurs for the smallest orders (highest wavelengths) and it is always greater than 20 pixels (15 μm pitch).
The 4C optical scheme adopted for the spectrograph is extremely compact and characterised by its planarity, i.e. the chief-ray, deviated by each optical element describes a plane. The light, coming from the dichroic and corrected by the ADC, is injected into the spectrograph from below such plane and redirected by a small folding mirror.
This type of planarity layout suggests considering a bench-like mechanical element, parallel to the plane defined by the chief-ray, as the reference element for the mechanical installation.
The NIR Spectrograph (Figure 6) is based on the same concepts of the UV-VIS one. Here the grating has a blaze angle of 44.0° and 72 grooves/mm. It is used at Littrow angle and with an off-plane angle of 5.0°. We plan to use the orders from 11 to 25. The selected glasses ($CaF_2$, Cleartran, S-FPL51Y) have been chosen to assure high throughput. The spectral format on the detector, that fits into a 2k x 2k 18 μm pixel detector, is shown in Figure 7.
Table 1 summarizes the main parameters of the spectrographs in the current baseline.

## 5.1 Acquisition and Guiding Camera

The acquisition of the target will be achieved with the Acquisition and Guiding Camera, whose optics must allow for a minimum field of view of 2 arcmin x 2 arcmin, a good transmission from 350 to 950 nm, and a good image quality (spot radius of 1 arcsec ~ 1 pixel). The camera will present a stop in the collimated beam to host filters.

Table 1. Spectrographs main parameters.

| Description | Unit | UV/VIS | NIR |
|---|---|---|---|
| **Input F/number** | | 11 | 11 |
| **Output F/number** | | 6.50 | 6.50 |
| **Slit height** | arcsec | 12.00 | 12.00 |
| **Collimator F/number** | | 6.50 | 6.50 |
| **Collimator Diameter** | mm | 45 | 45 |
| **Prism 1 Angle** | deg | 52 | 26 |
| **Prism 2 Angle** | deg | 52 | 26 |
| **Orders** | | 10 – 21 | 11 – 25 |
| **Blaze Angle** | deg | 41.770 | 44.000 |
| **Grooves/mm** | | 180.000 | 72.000 |
| **Groove spacing (Pitch)** | mm | 0.006 | 0.014 |
| **Off Blaze Angle** | deg | 0.0 | 0.0 |
| **Off Plane Angle** | deg | 3.35 | 5.0 |
| **Camera input F/#** | | 6.5 | 6.5 |
| **Camera output F/#** | | 3.6 | 3.47 |
| **Camera Diameter** | mm | 56 | 60 |
| **Angle on Sky per Pixel** | arcsec | 0.29 | 0.31 |
| **Camera Plate Scale** | arcsec/mm | 16 | 17 |
| **Detector Scale** | μm/arcsec | 61 | 59 |
| **Pixel Size** | μm | 15 | 18 |
| **No. of pixels A** | | 2048 | 2048 |
| **No. of pixels B** | | 4096 | 2048 |
| **Resolution for 1" slit** | | ~4600 | ~5000 |

### 5.2 Calibration unit

The purpose of the calibration unit is to provide light sources for flat fielding and wavelength calibration for the broad wavelength range of the two arms. The unit is divided into two arms: in one arm, the relatively bright lamps are mounted on an integrating sphere. This arm provides a very uniformly illuminated pupil at the cost of relatively poor efficiency due to loss from multiple reflections in the sphere. Another arm is used for faint lamps, where a diffusing plate is illuminated. Although more efficient, the exit distribution will be less homogeneous compared to the sphere. For each lamp, a relay lens is used to overlay the lamp exit diaphragm with the telescope pupil. A three-position mirror slide allows the user to select the light from each of the calibration sources or from the science field. The following lamps are foreseen on the integrating sphere: an UV-VIS flat field lamp (filtered QTH and/or Deuterium lamp); a NIR flat field lamp (QTH or TBD); a NIR spectral calibration lamp: Ar/Kr/Ne/Xe combination. In addition, an UV/VIS spectral calibration lamp (ThAr) will be mounted behind the diffuser.

### 5.3 Detector system

The detector for the UV-VIS arm will be a high performance, back illuminated CCD with a 15 μm square pixel, characterised for a high Quantum Efficiency (QE). The NIR detector will be an Infrared Array, according to the current baseline a 2048x2048 pixel hybrid array with a 18 μm pixel pitch (e.g. Teledyne H2RG). In the substrate-removed version, it can extend its operating wavelength range up to the visible band and increase the J band quantum efficiency. Moreover, a cut-off at 1.75 μm is available, then removing directly the K band from the detector.

The detector controller will be the ESO NGC (New Generation Controller) system. This controller can manage both CCDs and IR arrays. It is the current baseline for the ESO instrumentation and hence will be adopted in the SOXS project to comply with ESO standard and simplify maintenance.

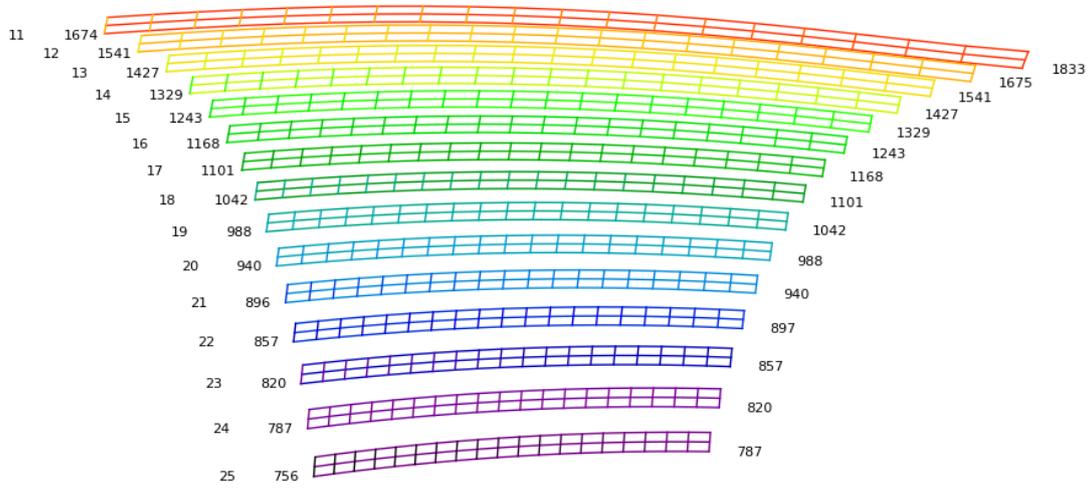

Figure 7. NIR spectral format.

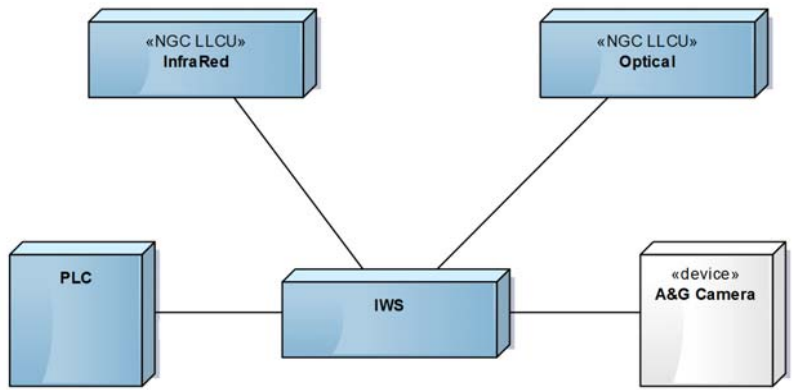

Figure 8. Control system concept.

## 6. CONTROL SYSTEM

The Instrumentation Software (INS) will be based on the VLT common software. In particular, it is assumed that the control system will be based, both in hardware and in software, on the latest COTS-based ESO standards for instrumentation control. This decision is justified by two considerations. First, all the requirements of the instruments in term of control and operations can be satisfied by an ESO standard control system. Second, this approach minimises risk, given the long experience present in the SOXS team on the development of control systems based on the ESO standards.

The SOXS Instrumentation Control System will consist of:
- one Instrument Workstation (IWS).
- one NGC LLCU for IR-DCS.
- one NGC LLCU for Optical-DCS.
- one GigE Vision device as Acquisition and Guide Camera.

- one or more PLCs for instrument functions control and sensor monitoring.

The SOXS software architecture will consist of:
- OS, Observation Software: coordinates the execution of exposures and interfaces to external sub-systems (telescope and archive). In particular, it provides for setup of the instrument functions and detectors, monitors the progress of exposures, collects science data, merges it with instrument status information and prepares the final files for archival.
- ICS, Instrument Control Software: controls all instrument functions, except detectors, and monitors all sensors.
- IR-DCS: carries out all tasks related to the control of the infrared detector sub-system. In particular, acquires detector data and transfers it to the IWS.
- Optical-DCS: carries out all tasks related to the control of the Optical detector sub-system. In particular, acquires detector data and transfers it to the IWS.
- Technical Camera control software: carries out all tasks related to the control of the Acquisition and Guiding detector sub-system. In particular, it acquires detector data and status information.

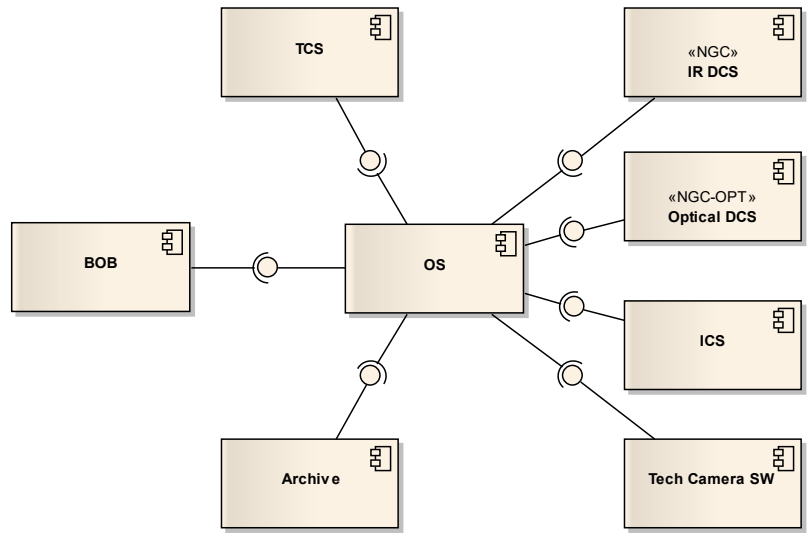

Figure 9. Instrument software architecture.

## 6.1 Devices Control

The SOXS instrumentation software will control the following instrument functions:
- Instrument shutter
- Calibration mirrors exchanger
- Acquisition and Guiding (A&G) slide
- One filter wheel for the A&G camera
- Two piezo actuators controllers
- One Atmospheric Dispersion Corrector (ADC)
- Two slit exchangers
- Visible camera focusing mechanism
- Calibration lamps

The SOXS instrumentation software will also monitor several sensors (temperature, pressure). It is assumed that the visible camera shutter will be controlled by the NGC and any shutter in the A&G camera will be controlled by the camera controller itself.

Concerning the piezo actuators, they will be used to keep the target in the slit compensating for flexures and differential refraction. They will be commanded through an analogue interface. The basic concept for their control is that look-up tables will be developed which will allow us to derive the positioning commands (analogue voltages) from the observing conditions. These commands will be then set by the instrument software in a digital to analogue output port of the control electronics.

## 7. OPERATIONS

The standard SOXS observing mode will be the simultaneous recording of spectra in the two UV-VIS and NIR arms in slit mode, using the dichroic to separate the beams. In slit mode the user can select, for each arm independently, a slit width among the available configurations (the minimum slit width will be approximately 0.5 arcsec).

Although the same object is observed in parallel by the two spectrographs, the integration times in the two arms can be different and multiple exposures can be taken in one arm while a single integration is carried out in the other.

The user can choose between templates for doing a single staring observation or a multiple observation with the target moved by a few arcseconds along the slit between the different exposures. For faint objects in spectral regions where sky subtraction is important, nodding will be adopted. The slits may be oriented at any parallactic angle using the NTT Nasmyth Instrument Rotator.

In the UV-VIS arm, where the atmospheric dispersion effect is more severe, an ADC is adopted in the design to correct for the effect in the corresponding wavelength range. The ADC is an integral part of the UV+VIS pre-slit optics, being always in the beam. The ADC control software is fed with altitude angle, temperature, pressure and relative humidity input values, which are used to compute the atmospheric dispersion and the angular position of the counter-rotating prisms.

The two spectrograph entrance slits have the same orientation and position on the sky. Once the orientation of the slit is set, the NTT Nasmyth adapter with the guide probe is locked on the guide star and the instrument follows the field rotation. The acquisition of the target will be achieved with the Acquisition and Guiding Camera, whose field of view is planned to be about 2 arcmin square. The camera system includes a filter wheel to isolate the different colour bands.

Each spectrographic arm has its own slits and the images of these slits in the focal plane must coincide with a reference pixel in the A&G unit.

After slewing the telescope to the target coordinates specified in the observing block, the target is identified by the instrument operator, the offset to move the reference wavelength to the position corresponding to the slit is automatically computed by the A&G software and applied to the telescope, taking into account the value of the reference wavelength and the central wavelength of the A&G filter in use. In the cases where the brightness of the targets is below the detection limit of the CCD (faint emission line objects, IR emitters), the acquisition will have to rely on offset centring, using a reasonably bright reference star in the field of view (blind offset).

Although the SOXS main observing mode is the slit spectroscopy, a simple imaging mode will also be implemented. The imaging mode adopts the Acquisition and Guiding camera and its set of filters. The acquisition images can be used to obtain reference photometry on both arms to flux calibrate spectra in addition to the usual spectrophotometric observations. Other applications can be the determination of magnitudes of transient objects such as GRB counterparts, supernovae, and variable objects.

## 8. CONCLUSIONS

SOXS is a single object wide-band spectroscopic facility for the ESO NTT dedicated to study transient phenomena, proposed by an international consortium and selected by ESO after a competitive down-selection phase in 2014-2015. The science case and the basics of the instrument proposal design have been presented. The design will evolve through the next phases of the project. The plan is to complete the instrument in a reasonable timescale of ~4 years, starting the operations immediately after its commissioning.